\input{psfig.sty}
\input{epsf.sty}

\def\mev{\,{\rm Me\kern-0.1em V}}
\def\gev{\,{\rm Ge\kern-0.1em V}}

\documentstyle[12pt]{article}

\begin{document}
\vspace*{-1.25in}
\small{
\begin{flushright}
FERMILAB-Pub-01/129-T \\[-.1in] 
June~2001 \\
\end{flushright}}
\vspace*{.75in}
\begin{center}
{\Large{\bf  Chiral Loops and Ghost States \\
in the Quenched Scalar Propagator}}\\
\vspace*{.45in}
{\large{W. ~Bardeen$^1$,
A.~Duncan$^2$, 
E.~Eichten$^1$, 
N. Isgur$^3$ and
H.~Thacker$^4$}} \\ 
\vspace*{.15in}
$^1$Fermilab, P.O. Box 500, Batavia, IL 60510 \\
$^2$Dept. of Physics and Astronomy, Univ. of Pittsburgh, 
Pittsburgh, PA 15260\\
$^3$Jefferson Lab, 12000 Jefferson Avenue, Newport News, VA 23606\\
$^4$Dept.of Physics, University of Virginia, Charlottesville, 
VA 22901
\end{center}
\vspace*{.3in}
\begin{abstract}
The scalar, isovector meson propagator is analyzed in 
quenched QCD, using the MQA pole-shifting ansatz to 
study the chiral limit.  In addition to the expected 
short-range exponential falloff characteristic of a heavy 
scalar meson, the propagator also exhibits a longer-range, 
negative metric contribution which becomes pronounced 
for smaller quark masses.  We show that this is a 
quenched chiral loop effect associated with the anomalous 
structure of the $\eta'$ propagator in quenched QCD.  
Both the time dependence and the quark mass dependence 
of this effect are well-described by a chiral loop diagram
corresponding to an $\eta'$-$\pi$ intermediate state, which  
is light and effectively of negative norm in the quenched approximation.   
The relevant parameters of the effective Lagrangian 
describing the scalar sector of the quenched theory are 
determined.

\end{abstract}

\section{Introduction}

For the foreseeable future, the quenched approximation 
will continue to play an important role in the study of lattice 
QCD for both practical and theoretical reasons.  The 
practical reasons are only too obvious, considering the 
orders-of-magnitude increase in computer resources 
required for comparable full QCD simulations.  But 
quenched QCD also provides a very instructive counterpoint  
to the full theory which can yield useful insight into the 
effect of quark loops in determining hadron structure. 
This is particularly true in the light-quark limit, where 
lattice results can be interpreted theoretically with 
the aid of an effective field theory corresponding to 
quenched chiral perturbation theory.

The introduction of quenched chiral perturbation 
theory by Sharpe, Bernard, and Goltermann 
\cite{Sharpe,B&G} established the framework for 
analyzing the chiral behavior of quenched QCD. This 
work focused attention on the particular, anomalous 
role of the quenched $\eta '$ propagator. The physical  
$\eta'$ is approximately a flavor singlet and gets most of its mass from the axial anomaly via repeated  $q\bar{q}$ annihilation.
In the quenched approximation, a single annihilation 
vertex can appear if both sides of the ``hairpin diagram'' 
are attached to valence quark lines, but repeated 
annihilation cannot take place without the closed quark 
loops of the unquenched theory.  Instead of cancelling 
the Goldstone pole in the valence quark (``connected'') 
diagram, the single annihilation diagram contributes a 
term to the $\eta '$ propagator which has a {\it double} 
Goldstone pole and an overall sign opposite to that of 
the valence diagram. 
Unlike full QCD where $\eta '$ loops remain infrared finite 
in the chiral limit, in the quenched theory, the double 
Goldstone pole of the hairpin term produces additional 
quenched chiral loop (QCL) singularities which alter the 
chiral behavior of the theory. This QCL effect was first 
observed in lattice results as a deviation from linear 
behavior of the squared pion mass as a function of 
quark mass \cite{lat98,CPPACS}. Recently, the 
quenched chiral limit was extensively investigated in a 
high-statistics study at $\beta=5.7$\cite{chlogs}.
The problem of exceptional configurations, which had 
prevented previous studies from probing the smaller 
quark mass region of $m_{\pi}<350$ MeV, was resolved 
by the pole-shifting ansatz of the modified quenched 
approximation (MQA)\cite{MQA}. The QCL effect was 
observed in both the pseudoscalar mass and decay 
constants at the level expected from a direct study 
of the $\eta '$ hairpin mass insertion and the topological 
susceptibility. This established several independent and 
quantitatively consistent determinations of the 
quenched chiral log coefficient parameter $\delta$ (or,
equivalently, of the $\eta'$ mass insertion $m_0^2$).

The spectroscopy of scalar mesons has long been 
one of the murkier areas of hadron phenomenology. 
The well-established scalar mesons are typically at 
masses of a GeV or more, although there have been 
occasional, fleeting experimental indications of lower 
mass scalar resonances. In the flavor singlet sector, 
mixing with glueball states further complicates the 
intepretation of experimental data.  This is clearly an 
area where lattice QCD calculations can be expected 
to play a crucial role in the future. High-statistics 
quenched studies will undoubtedly be an important 
part of this effort. Naively, we might expect that the 
anomalous chiral behavior induced by the quenched 
approximation would be of little importance for 
scalar-meson spectroscopy, since the meson masses 
involved are all expected to be quite heavy and not 
particularly sensitive to the details of chiral extrapolation.  
However, as we show in this paper, a more careful 
theoretical analysis combined with a precise numerical 
study of the scalar, isovector meson propagator at 
very light quark mass reveals perhaps the most striking 
quenched chiral loop effect yet observed.  

\begin{figure}
\vspace*{2.0cm}
\includegraphics{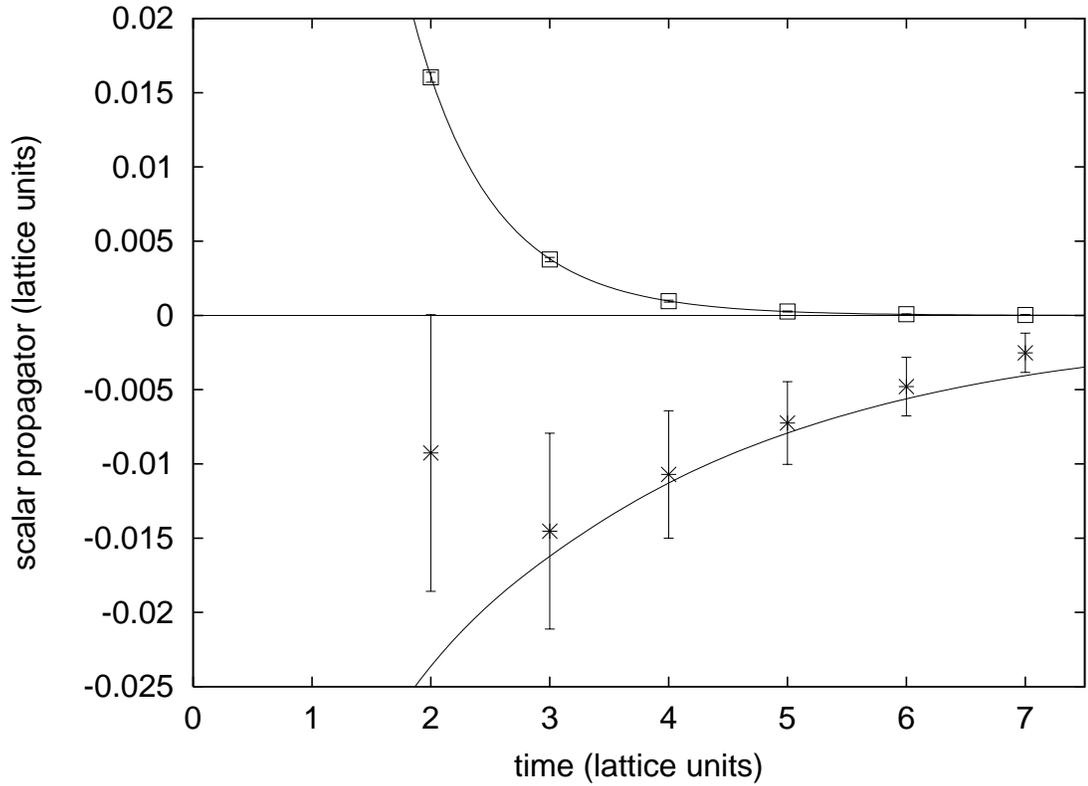}
\vspace{11.5cm}
\caption[]{The scalar propagator for the heaviest ($\kappa
=.1400$) and lightest ($\kappa=.1428$) quark masses studied. For 
heavy quarks (boxes) the propagator is positive and exponentially
falling. The (upper) solid curve is a two-exponential fit.
For light quarks (*'s) the propagator exhibits a negative quenched chiral
loop effect.} The lower solid curve is the prediction of one-loop chiral perturbation
theory for this term. (See Section 2.)
\label{fig:sc_heavylight}
\end{figure}

\begin{figure}
\vspace*{2.0cm}
\includegraphics{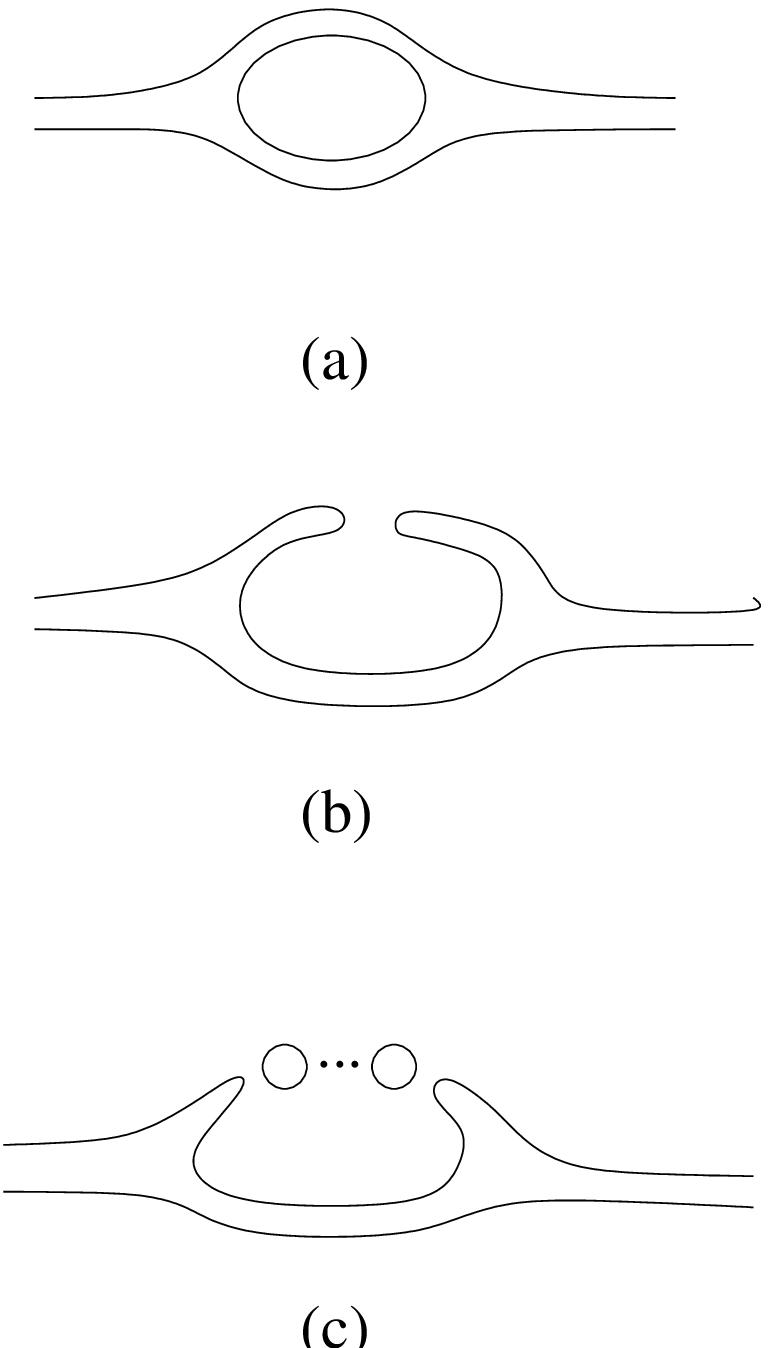}
\vspace{11.5cm}
\caption[]{Quark line diagrams which contribute to the $\eta '-\pi$ intermediate
state in the scalar isovector propagator.  }
\label{fig:quark_lines}
\end{figure}

The chiral loop effect we discuss here is exhibited very 
clearly by comparing the scalar, isovector propagator for 
the heaviest and lightest quark
masses studied. These are shown in Figure \ref{fig:sc_heavylight}. 
For the heaviest quark mass the propagator is a positive and 
rapidly falling exponential, as expected from 
heavy scalar meson intermediate states. The fit shown is a sum of two exponentials 
with the excited state $a_0^*$ mass extracted by comparing local-local, 
smeared-local, and smeared-smeared correlators. (See discussion in Section 3.) 
By contrast, the scalar propagator for lighter quarks 
exhibits a qualitatively different behavior. 
[The lower curve in Figure \ref{fig:sc_heavylight} is the one-loop quenched 
$\chi PT$ prediction for the contribution to the scalar propagator of the 
$\eta'$-$\pi$ intermediate state only. See below and Section 2.]
This unusual behavior
has two qualitative features which point 
clearly to a specific theoretical interpretation: 
(1) The additional component becomes much more 
prominent for the lightest quark masses, 
and (2) the sign of this component is negative (i.e. 
opposite to that required by positivity of the spectral 
function). Both of these properties suggest that the new 
component is an effect of the $\eta'$-$\pi$ 
intermediate state, which is light in the quenched 
approximation.  This contribution arises from the
``hairpin + pion'' diagram shown in  
Figure \ref{fig:quark_lines}(b). This interpretation 
also explains the fact that this component 
has a negative spectral weight. The hairpin (single 
mass insertion) diagram appears in the $\eta'$ 
propagator with the opposite sign from that of the 
full propagator.  In full QCD the $\eta'$-$\pi$ intermediate 
state would arise from three different types of quark 
line diagrams, Figure \ref{fig:quark_lines}(a) with a 
single vacuum loop going around the entire diagram, 
Figure \ref{fig:quark_lines}(b) with a single hairpin 
vertex, and the graphs of type \ref{fig:quark_lines}(c) with one or more 
vacuum bubbles inserted into the $\eta '$ propagator.  
Of these three, only the hairpin diagram, 
Figure \ref{fig:quark_lines}(b), is included in the 
quenched approximation. Since the single hairpin insertion has an overall negative sign 
(corresponding to a positive $\eta'$ mass shift), the sign 
of this contribution is opposite to that which would be required by 
spectral positivity. The prominence of this negative term in our measured 
propagator is a clear example of the unitarity 
violation induced by the quenched approximation. [In the Bernard-Goltermann 
scalar ghost quark formulation of quenched chiral perturbation theory \cite{B&G},
the negative sign of the scalar propagator arises from the dominance of a
negative metric state consisting of a pair of ghost-quark mesons.]

\section{Lattice Results for the Scalar Propagator}

In a recent study of quenched chiral logs and the 
$\eta '$ propagator \cite{chlogs}, the MQA pole-shifting 
ansatz was applied to a set of quark propagators 
for an ensemble of 300 gauge configurations at 
$\beta=5.7$ on a $12^3\times 24$ lattice. The 
fermion action was clover-improved Wilson-Dirac 
with $C_{sw}=1.57$. Nine values of quark mass 
were used, corresponding to a range of hopping 
parameters from $\kappa = .1400$ to .1428. 
Valence quark propagators were calculated from both 
local delta-function sources and from exponentially 
smeared sources in Coulomb gauge. (For further details, see 
Reference \cite{chlogs}.) These pole-shifted quark 
propagators were used to calculate the scalar meson 
propagators considered in this paper. The pion 
masses obtained from this ensemble have values 
ranging from $m_{\pi}a=0.245$ to $0.603$, and are listed
in Table I.

\begin{figure}
\vspace*{2.0cm}
\includegraphics{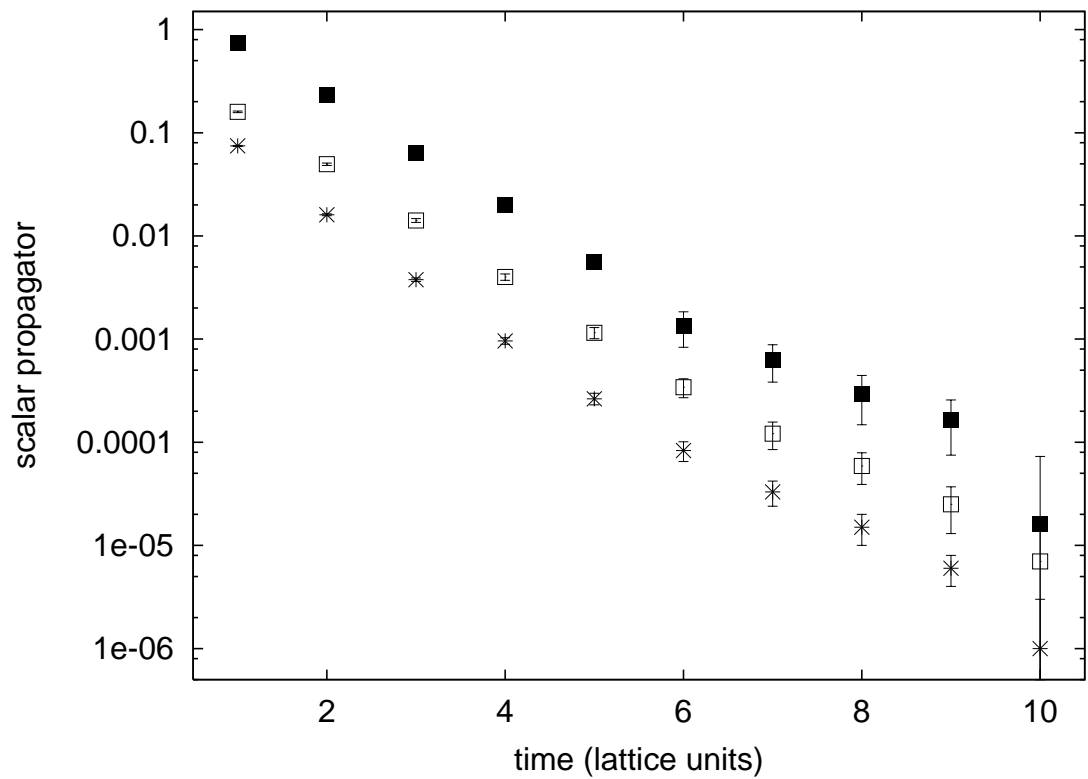}
\vspace{11.5cm}
\caption[]{The scalar propagator for the heaviest quark ($\kappa
=.1400$) with local-local (*'s), smeared-local (empty boxes), and smeared-smeared (full boxes)
sources.}
\label{fig:sc1400_logplot}
\end{figure}

\begin{figure}
\vspace*{2.0cm}
\includegraphics{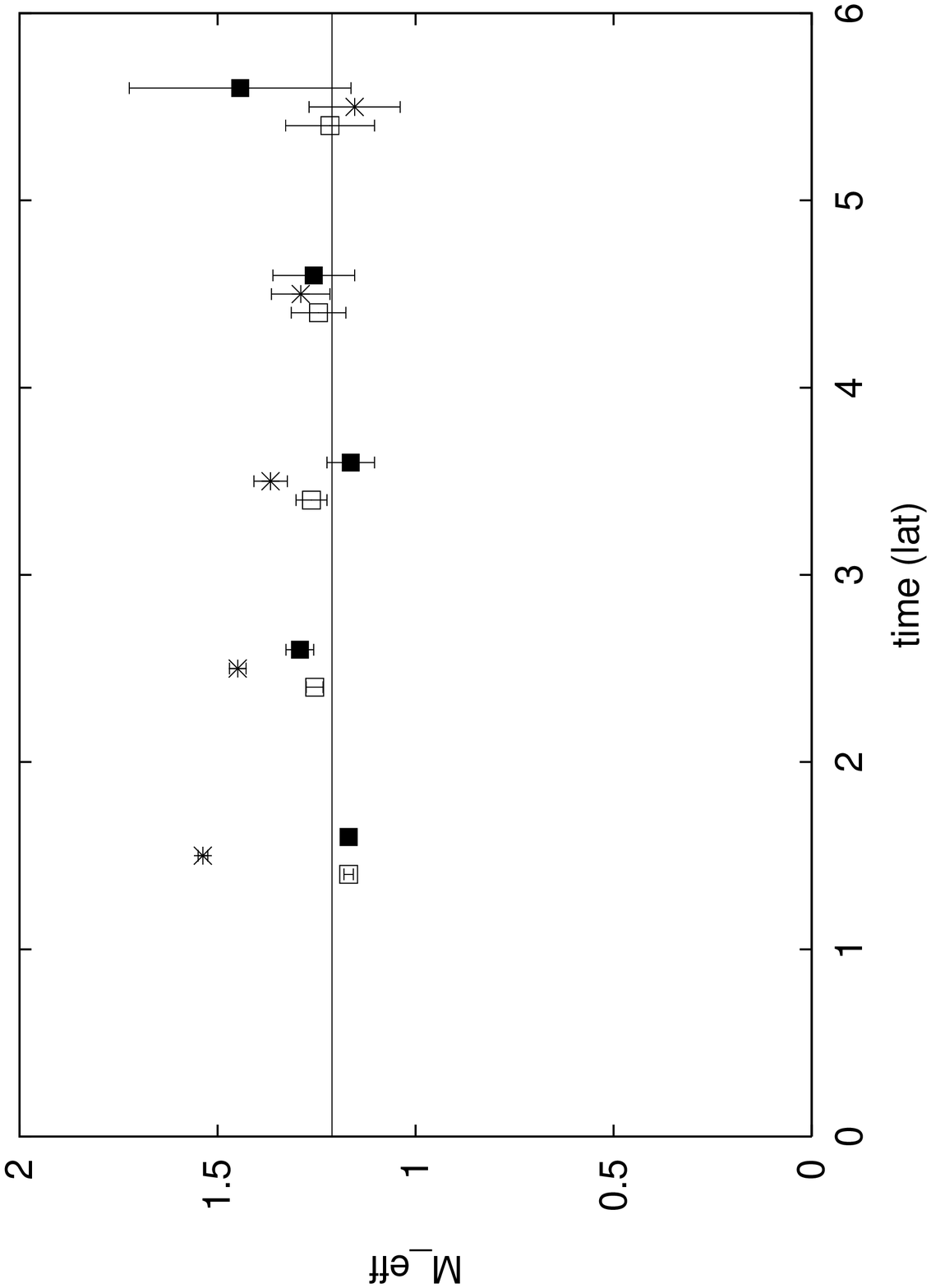}
\vspace{11.5cm}
\caption[]{Effective mass plot for $\kappa = .1400$ for local-local (*'s),
smeared-local (empty boxes), and smeared-smeared (full boxes) correlators.  }
\label{fig:effective_mass}
\end{figure}

It is natural to expect that the scalar $\bar{\psi}\psi$ correlator in QCD will 
be dominated by the coupling to the lightest scalar 
meson which is expected to have a mass larger than 
$1$ GeV. For the heaviest quark masses  studied here, this is in fact the behavior observed for the quenched propagator. First, consider the scalar propagator for our heaviest quark mass, with 
$m_{\pi}a=.603,$ corresponding to 
$\kappa=.1400$. (From Reference \cite{chlogs}, we have $\kappa_c = 0.14329$
for clover-improved Wilson-Dirac fermions).  
The measured values of the local-local, smeared-local, and smeared-smeared propagators 
are shown in the log plot in
Figure \ref{fig:sc1400_logplot}. As expected, the plot shows
clear evidence of a massive scalar meson.   For
the smeared-local propagator, the time-dependence is 
reasonably well described for $t>2$ by an exponential 
fit with a mass of $M_sa=1.25(2)$. 
The effective mass plot for the smeared-local propagator 
is shown in Figure \ref{fig:effective_mass} for 
$\kappa=.1400$. For comparison, the effective mass 
for the local-local and smeared-smeared propagators are 
also plotted.  Although the smearing function used (an exponential with 
exponent $0.5$ in lattice units) was not tuned for this 
particular problem, it does a good job of 
removing excited state contamination, giving a reasonably 
flat effective mass over several time slices from 
$t=2$ to $t=6$, and a nearly identical effective mass plot for smeared-local and smeared-smeared propagators (indicating an absence of excited states in both).
  
\begin{figure}
\vspace*{2.0cm}
\includegraphics{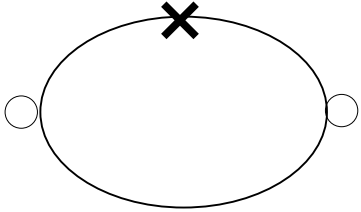}
\vspace{1.5cm}
\caption[]{One-loop quenched chiral perturbation theory graph evaluated in Eq. (\ref{eq:B}). }
\label{fig:bubble}
\end{figure}

\begin{figure}
\vspace*{2.0cm}
\includegraphics{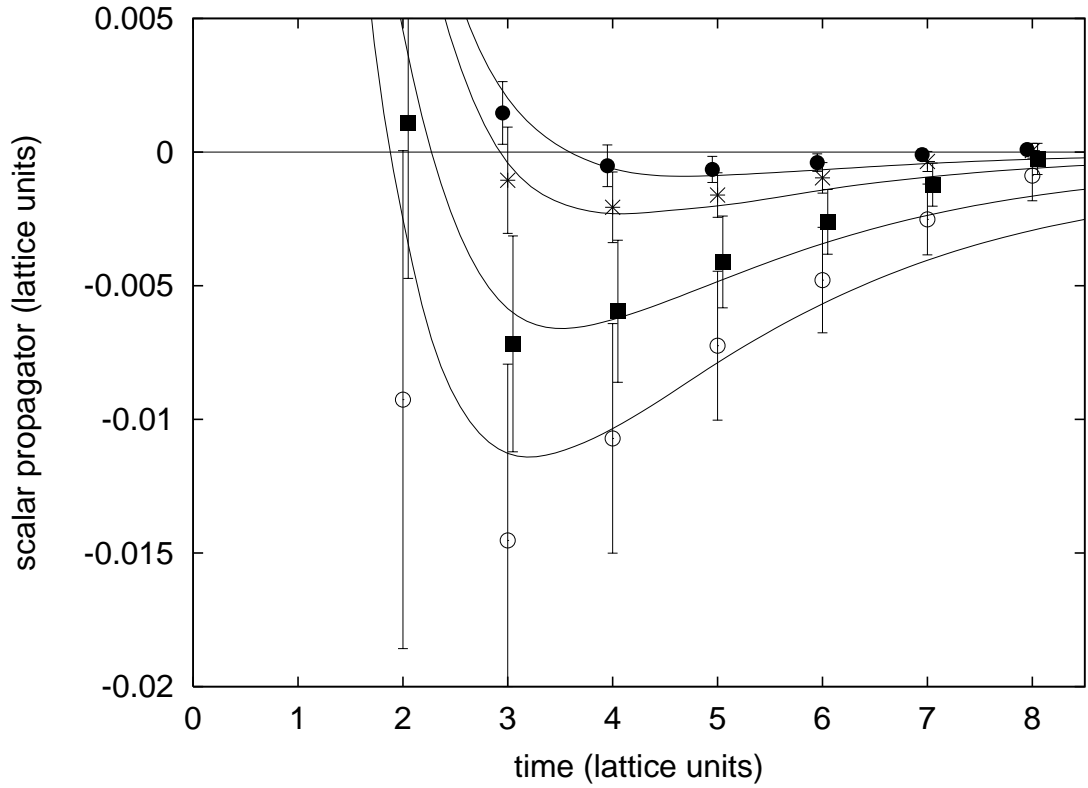}
\vspace{11.5cm}
\caption[]{ Scalar propagator for $\kappa = .1423 (\bullet)$, .1425 (*'s), 
.1427 (solid squares), and .1428 $(\circ).$  
Solid curves are fits to the bubble sum formula, (\ref{eq:bubblesum}).}
\label{fig:bsfits}
\end{figure}

As pointed out in the Introduction, the behavior of the scalar propagator changes dramatically for the lighter quark 
mass values.  The results for the scalar propagator for 
our lightest quark mass values are shown in 
Figure \ref{fig:bsfits}, with  $m_{\pi}a=.245,.267,.307,$ and $.342$ 
corresponding to $\kappa$ values  ($.1428,.1427,.1425$, and .1423) 
respectively.  Here we have plotted the local-local $\bar{\psi}\psi$  
correlators.   Instead of exhibiting the behavior 
expected for scalar meson states, these propagators 
are dominated by a significant negative contribution 
in the range $t=2$ to $t=7$ which increases for lighter 
quark masses.  We interpret this behavior as a clear 
signal for the negative metric contribution associated with
the $\eta '$-$\pi$ intermediate state discussed in 
the previous Section.  

\begin{figure}
\psfig{figure=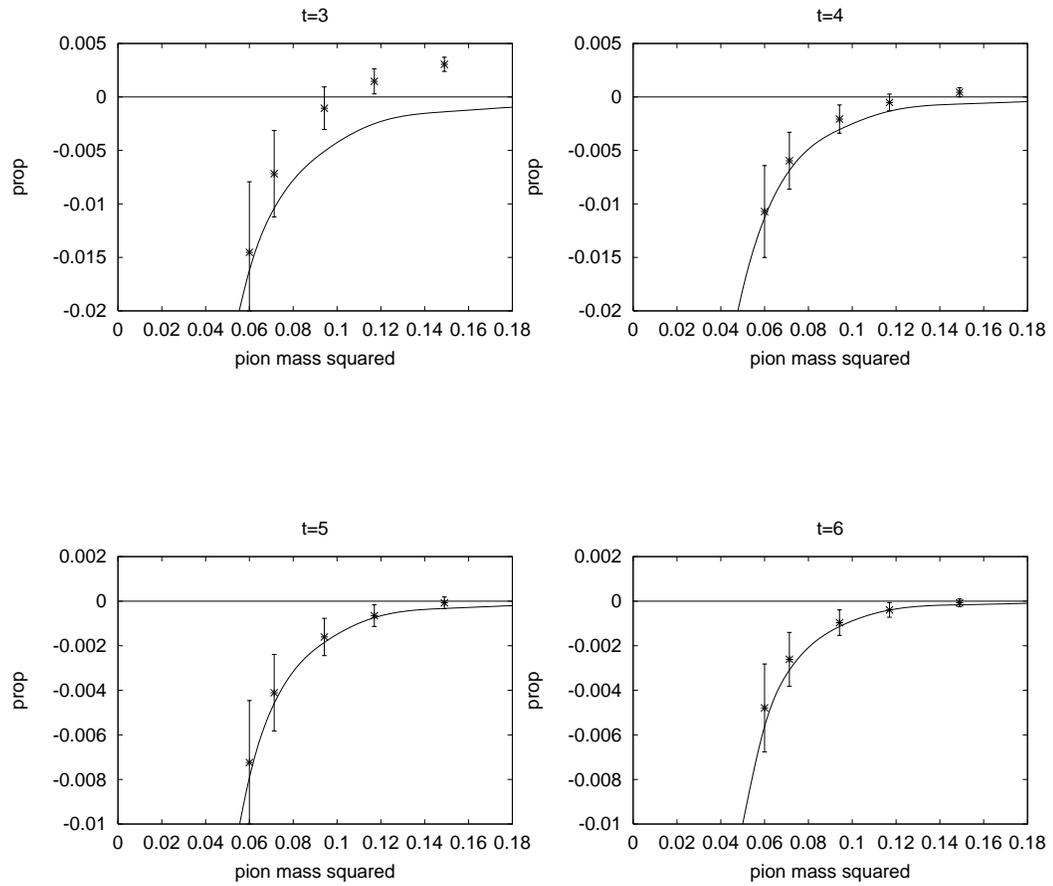,width=1.00\hsize}
\vspace{-1in}
\caption[]{ Pion mass dependence of the scalar propagator for $t=3,4,5,6$.
}
\label{fig:t3456}
\end{figure}

The $\eta'$-$\pi$ loop interpretation of the anomalous component of the scalar propagator
is made more convincing by noting that the leading 
chiral behavior of this contribution is
entirely determined by current algebra in terms of parameters 
which have already been measured with this ensemble in our previous 
study of the pseudoscalar propagator \cite{chlogs}.
The relevant chiral loop diagram is shown in Figure \ref{fig:bubble}, with the $\times$
denoting the $\eta'$-hairpin mass insertion $m_0^2$, and the $\circ$'s representing
$\bar{\psi}\psi$ operator vertices. If we ignore form factor effects at these
vertices, we can calculate the $\eta'$-$\pi$ contribution to the $\bar{\psi}\psi$ propagator in momentum space. The propagator 
\begin{equation}
\Delta(p) \equiv \sum_{x} e^{-ip\cdot x}\langle\bar{\psi}\psi(x)\bar{\psi}\psi(0)\rangle
\end{equation}
is given in terms of the one-loop integral $B(p)$,
\begin{equation}
\label{eq:B}
\Delta(p) \sim -4r_0^2 B(p)
\end{equation}
where
\begin{eqnarray}
\label{eq:Bformula}
B(p) &=& {1\over VT} \sum_{k} {1\over ((k+p)^2 + m_{\pi}^2)} 
      {m_0^2\over (k^2+m_{\pi}^2)^2 }
\end{eqnarray}
Here, $2r_0$ is the $\bar{\psi}\psi$ to $\eta'$-$\pi$ coupling. (In the next section 
we will discuss including the scalar meson intermediate state, along with form-factor/unitarization effects).
The pion mass values for the hopping parameters used here have already been
reported in Reference \cite{chlogs} and are listed here in Table I.
The hairpin insertion mass $m_0^2$ is also determined in Reference \cite{chlogs}. For 
clover improved fermions with $C_{sw}=1.57$ the value found in the chiral limit 
(dropping a flavor factor of $\sqrt{3}$ included in Table II of that reference) is 
\begin{equation}
\label{eq:m0}
m_0 = .33(2)
\end{equation}
The quark mass dependence of $m_0$ was found to be very mild (see Table II of 
Reference \cite{chlogs}). Here, for simplicity, we will take it to be a mass 
independent constant given by (\ref{eq:m0}). The only other parameter needed to 
evaluate the $\eta'$-$\pi$ bubble
contribution is the coupling of the scalar $\bar{\psi}\psi$ operator to the 
$\eta'$-$\pi$ state,
\begin{equation}
\label{eq:scdensity}
\langle 0|\bar{\psi}\psi|\eta'\pi\rangle
\end{equation}
In the quenched theory in the chiral limit, we can relate the threshold value of this matrix element, via a soft-$\eta'$ reduction, to the decay constant $f_P$, the 
vacuum-to-one-pion matrix element of the pseudoscalar density. By this argument, 
we relate the coupling constant in ($\ref{eq:B}$) to the $m_{\pi}^2$ vs. $m_q$ slope parameter $r_0$ 
determined in Reference \cite{chlogs},
\begin{equation}
\label{eq:r0}
r_0 = 1.99(12)
\end{equation}
Thus, in this approximation the effect of the $\eta'$-$\pi$ bubble, 
Eq. (\ref{eq:B}), is completely predicted by chiral symmetry in terms of the 
constants (\ref{eq:m0}), (\ref{eq:r0}), and the pion masses in Table I.
The result for the lightest quark mass ($\kappa = .1428$) is plotted along 
with the measured propagator in Figure \ref{fig:sc_heavylight} (lower curve). 
Considering that the plotted curve has 
{\it no adjustable parameters}, it fits the size and shape of the propagator 
remarkably well in the time range from $t=3$ to $t=7$. For times $t<3$, the 
measured propagator deviates from the one-loop expression in the positive 
direction, indicating that the spectral function also contains a heavy scalar 
meson. This short-range, positive metric exponential component dominates 
in the heavier quark cases where the chiral loop effect becomes unimportant. 
The one-loop result also explains the 
pion mass dependence of the anomalous propagator component, as shown in 
Figure \ref{fig:t3456}, where we have plotted the value of the scalar propagator at 
various time separations as a function of $m_{\pi}^2$. The solid curves 
are the $p_0$ Fourier transform at $\vec{p}=0$ of the one-bubble 
integral (\ref{eq:B}), with $m_0$ and $r_0$ given by (\ref{eq:m0}) 
and (\ref{eq:r0}). Again the agreement with the data is surprisingly good, 
considering the absence of any adjustable parameters. The accurate description of both the 
time-dependence and the pion mass dependence leaves little doubt that, for 
the lightest quark masses, the scalar propagator over the range of times 
from about $t=3$ to 7 is completely dominated by the $\eta'$-$\pi$ loop. 
Furthermore, this agreement shows that the soft pion theorem relating the 
scalar-density matrix element (\ref{eq:scdensity}) to the corresponding pseudoscalar matrix 
element is  well satisfied by the lattice results. This is somewhat surprising at $\beta=5.7$, where one might have expected substantial chiral symmetry violation from finite lattice-spacing effects.

For short time separations and/or heavier quark masses, an accurate description 
of the scalar propagator requires positive contributions from scalar meson 
states along with the negative $\eta'$-$\pi$ component.
In the next Section, we present a derivation of the quenched scalar 
propagator based on a detailed chiral Lagrangian 
description which will be used to fit our data for 
all quark mass values. The corresponding propagator fits are
discussed in Section 4.

\section{Quenched Chiral Perturbation Theory}

According to the analysis of Bernard and 
Golterman, low-energy quenched QCD is described by an effective 
local field theory with degrees of freedom corresponding 
to meson bound-states, including not only ordinary $q\bar{q}$ mesons, 
but also $q'\bar{q}$, $q\bar{q}'$, and $q'\bar{q}'$ mesons, where 
$q'$ is a wrong-statistics ghost quark. Order by order in the $\eta'$ 
mass insertion $m_0^2$, the ghost-quark formalism is easily seen to be equivalent to Sharpe's formulation, which begins with a $U(3)\times U(3)$ chiral
Lagrangian combined with rules inferred from the structure of quark-line
diagrams in the quenched theory.  In this paper, as in our previous analysis \cite{chlogs}
we follow the latter approach, invoking a chiral Lagrangian having valence pion and 
$\eta '$ degrees of freedom.  To describe quenched 
QCD, the chiral Lagrangian is supplemented with 
rules which reflect the suppression of internal quark 
loops.  As a result, all normal pion loops are suppressed 
except for cactus diagrams involving the $\eta '$ meson
connected by single insertions of the hairpin mass term.  
Using this procedure we obtained a consistent fit to our 
lattice data for the pseudoscalar valence and $\eta '$ hairpin propagators 
and determined the relevant chiral Lagrangian parameters.   
We now extend this analysis to the isovector, scalar 
propagator.

The appropriate effective field theory must now include 
a multiplet of scalar mesons in addition to the pion and 
$\eta '$ degrees of freedom.   Using standard chiral 
Lagrangian methods, the heavy scalar mesons are 
described by scalar fields transforming nonlinearly 
under chiral symmetry rotations.  The resulting chiral 
Lagrangian is
\begin{eqnarray}
{\cal L} &=& {f^2 \over 4} tr \{ \partial U \partial U^{\dagger} \} 
              +  {f^2 \over 4} tr \{ \chi^{\dagger} U 
                    + U^{\dagger} \chi  \}   \nonumber \\
         &+&  {1 \over 4} tr\{ D\sigma D\sigma \} 
             - {1 \over 4} m^2_s tr\{\sigma \sigma \} 
           + g_s tr\{\sigma \surd{U}\partial U^{\dagger}\partial 
                   U\surd{U^{\dagger}} \} \nonumber \\ 
         &+& f_s tr\{\chi^{\dagger}\surd{U}\sigma \surd{U} 
                     +\chi \surd{U^{\dagger}}\sigma \surd{U^{\dagger}} \} 
              + {\cal L}_{\rm hairpin} 
\end{eqnarray}
where $\cal D$ is a chiral covariant derivative, 
$m_s$ the sigma mass, $g_s$ the strength of 
the chiral invariant coupling of sigma mesons to 
pions and $f_s$ the strength of the corresponding chiral symmetry 
breaking interactions. Here we denote the scalar field by $\sigma$. (Elsewhere, when referring to particle states, we use the conventional $a_0$ and
$a_0^*$ to denote ground-state and excited-state mesons.) 
The chiral field matrix 
$U$ contains the pion and $\eta '$ degrees of 
freedom.  The $\eta '$ mass term is given by the 
hairpin Lagrangian,
\begin{equation}
{\cal L}_{\rm hairpin} = -{1\over 2} m_0^2 (f^2/8) 
     [i tr\ln (U^{\dagger}) - i tr\ln (U)]^2 . \;
\end{equation}
We have not included interactions involving the 
couplings $L_5$ and $L_8$ used in our previous 
analysis of higher order terms in the sigma potential 
as they will have only small effects on our analysis 
of the isovector scalar propagator.   

Scalar and pseudoscalar quark densities are 
represented by meson operators in the effective 
field theory and can be determined from the 
dependence of the chiral Lagrangian on the spurion 
field $\chi$.   Hence the isovector scalar density 
operator is given by
\begin{eqnarray}
\overline{\Psi_2} \Psi_1 &=& - {1 \over 2} r_0 f^2  (U + U^{\dagger})_{12}
  \nonumber \\
     &&  - 2 r_0 f_s  (\surd{U}\sigma\surd{U}  
	+ \surd{U^{\dagger}}\sigma\surd{U^{\dagger}})_{12} \;
\end{eqnarray}
with
\begin{eqnarray}
 \chi = \chi^{\dagger} = m_{\pi}^2 = 2 r_0 m_{\rm quark} . \;
\end{eqnarray}
Our previous fits to the pseudoscalar propagator 
determined the following values for the chiral 
Lagrangian parameters,
\begin{eqnarray}
\label{eq:fixedparams}
  f  &=& 0.1066(24) \nonumber \\
 m_o &=& 0.33(2) \nonumber \\
 r_o &=& 1.99(12).  \;
\end{eqnarray}

To lowest order, the local isovector scalar density 
may be expanded in terms of the pion, $\eta '$ and 
sigma fields,
\begin{equation}
\overline{\Psi_2} \Psi_1 = 2 r_0 (\pi^+\eta ') - 4\sqrt{2} r_0 f_s (\sigma^+). \;
\end{equation}
At tree level, only the sigma propagator gives 
a positive contribution to the scalar propagator
\begin{equation}
<\overline{\Psi_2} \Psi_1~~ \overline{\Psi_1} \Psi_2 > = 32 r_0^2 f_s^2 P_{\sigma} . \;
\end{equation}
where $P_{\sigma}$ is the $a_0$ meson propagator.
The $\eta '$-$\pi$ term can only contribute via a meson 
loop.  However, the $\eta '$ propagator in the loop 
is the $\eta '$ hairpin propagator with a single insertion 
of the hairpin mass, $m_0^2$ (see Figure \ref{fig:quark_lines}).   Including only these 
terms, the scalar propagator is
\begin{equation}
< \overline{\Psi_2} \Psi_1~~ \overline{\Psi_1} \Psi_2 > = 32 r_0^2 f_s^2 P_{\sigma} 
                    + 4 r_0^2 B_{\rm hp} \;
\end{equation}
Here  $B_{hp}$ is the
hairpin bubble
\begin{equation}
B_{\rm hp} = FT \{ \frac{1}{VT}\sum_{k} {1 \over ((k+p)^2 + m_{\pi}^2)} 
      {-m_0^2 \over (k^2+m_{\pi}^2)^2 } \} \;
\end{equation}
and $FT$ means the Fourier transform on $p$. As discussed in the
Introduction, the negative 
sign of the $\eta '$ propagator in loops is dictated by 
the quenching and implies that this meson loop makes 
a negative contribution to the scalar propagator.   It 
should also be noted that the $\eta '$ propagator is 
more infrared singular than the normal valence 
pion propagator.  This has a dramatic effect on the 
scalar propagator.   In the light quark limit, the quenched 
theory predicts that large negative chiral loop effects 
should dominate the scalar propagator.   Our lattice 
data give convincing evidence for these effects.

A complete analysis of the scalar propagator must 
go beyond the lowest order terms given above.   The 
chiral Lagrangian predicts couplings between the scalar meson
and $\eta '$-$\pi$ states as well as $\eta '$-$\pi$ 
rescattering interactions generated by the symmetry 
breaking terms that give mass to the pion. These 
higher order terms involve $\eta '$-$\pi$ bubble 
diagrams and can be resummed into a closed form.    
Since the chiral invariant interactions between the 
$\sigma$ and $\eta'$-$\pi$ states include derivative couplings, 
bubbles with derivative couplings at the $\sigma\eta'\pi$ 
vertex must, in principle, also be 
included. These terms are less infrared singular than
the nonderivative terms and are not expected 
to play a major role in explaining the large quenched 
effects seen in the data.  In the following analysis, we will neglect 
such derivative coupling terms. The formulas and results given below are for the 
equal quark mass case; generalizations to the unequal 
mass case are straightforward. 

\begin{figure}
\vspace*{2.0cm}
\includegraphics{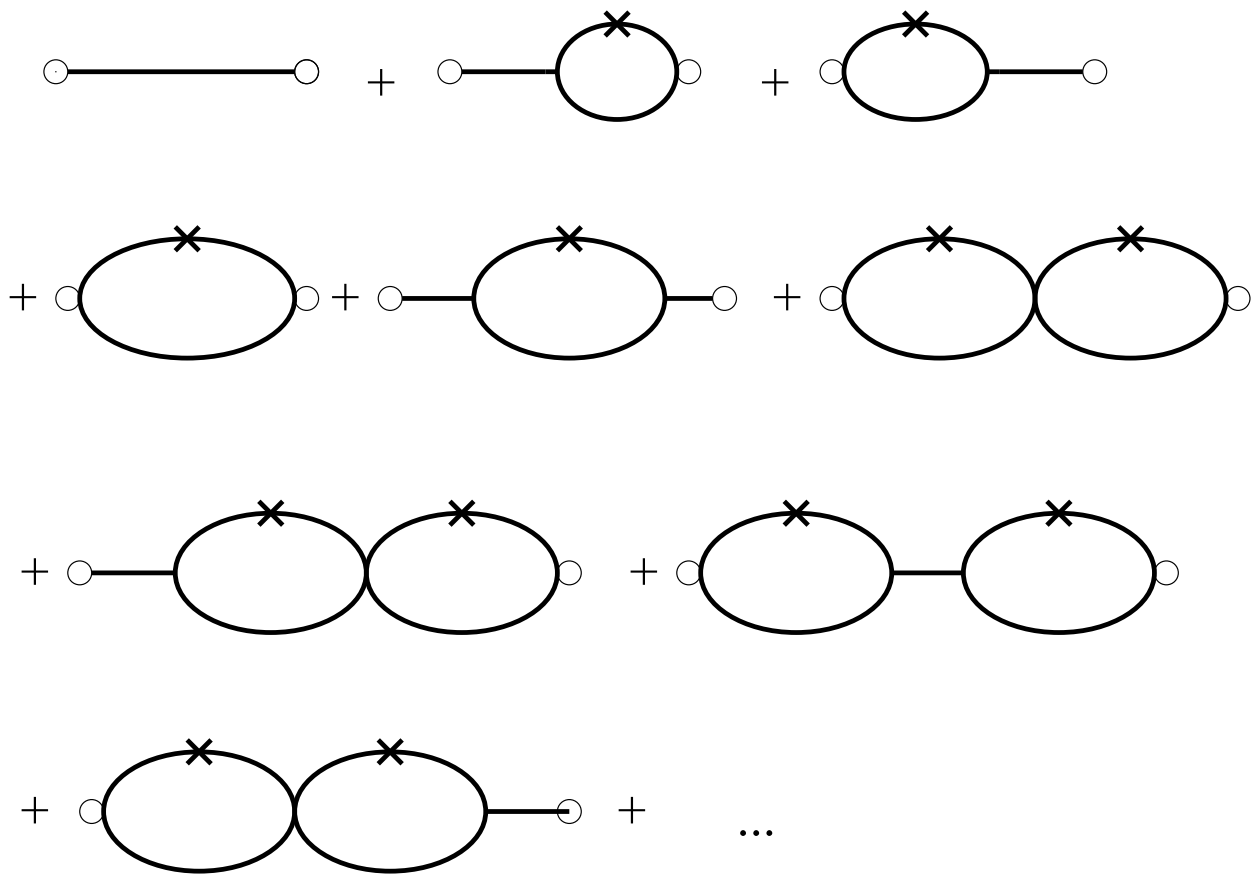}
\vspace{11.5cm}
\caption[]{Graphs which are included in the resummed scalar propagator, Eq. (\ref{eq:bubblesum}).
.  }
\label{fig:bubblesum}
\end{figure}

As is always the case in propagator studies, the presence of excited states in the spectral function complicates the analysis. 
As we discussed in Section II, in addition to the propagator for local
$\bar{\psi}\psi$ operators, we have calculated propagators 
with a smeared source at one or both ends. In our previous analysis of the 
pseudoscalar channel \cite{chlogs}, a full multi-state 
fitting procedure was employed. Here we take a somewhat simpler approach which is suggested by features of the 
data. We model the spectral function in terms of three components: 
(1) the $\eta'$-$\pi$ state, (2) a ground-state scalar meson $a_0$, and 
(3) an excited-state scalar meson $a_0^* $. We perform
an analysis of the local-local propagator by first using information from the smeared-local
and smeared-smeared propagators to 
remove the excited-state $a_0^*$ scalar meson component and then fitting to a formula obtained by resumming 
all repeated bubble graphs involving the $\eta'$-$\pi$ state and the ground-state $a_0$ scalar meson, as 
depicted in Figure \ref{fig:bubblesum}. 
To estimate and remove the contribution of the excited $a_0^*$ from the local 
propagator, we consider first the case with the heaviest quark, $\kappa=.1400. $
Within statistics, all of the scalar propagators for smeared and local sources exhibit the same 
time-dependence beyond $t=3$. In fact, the smeared-local and smeared-smeared 
propagators are equal, up to an overall factor, at all time slices, as evidenced by the
effective mass plot in Figure \ref{fig:effective_mass}. 
For the heavier quark mass values, the $\eta'$-$\pi$ loop is relatively 
unimportant, and the smeared and local propagators can be analyzed in terms of a ground-state 
and excited-state scalar meson. The fact that the smeared-local and smeared-smeared propagators, 
$\Delta_{SL}(t)$ and $\Delta_{SS}(t)$ are proportional and that both give flat effective mass plots 
indicates that neither propagator has a significant excited-state component. On the other hand, 
the local-local propagator clearly has a significant $a_0^*$ component, as seen in the effective 
mass plot, Figure \ref{fig:effective_mass}. 

It is straightforward to estimate 
the $a_0^*$ contribution by using the (essentially time-independent) ratio 
$\Delta_{SL}/\Delta_{SS}$ to determine the excited state component of $\Delta_{LL}$. 
\begin{equation}
\Delta_{LL}(t) - \frac{\Delta_{SL}}{\Delta_{SS}}\Delta_{SL}(t)\sim
C_{a_0*}\exp(-m_{a_0*}t)
\end{equation}
Numerically, the above procedure applied to the $\kappa=.1400$ propagators gives excited state 
parameters 
\begin{equation}
\label{eq:heavy}
C_{a_0^*} = 0.275(30)\;,\;\;\;\; m_{a_0^*} = 1.86(10)
\end{equation}
The large mass obtained emphasizes that the excited state contribution is only important at very short 
time separations. Similar results are obtained for $\kappa=.1405$ and $.1410$, with no significant 
observed dependence on quark mass. 
With this information, we can separate off the excited state 
component of the local-local propagator:
\begin{equation}
\Delta_{LL}(t) = C_{a_0^*}\exp(-m_{a_0^*}t) + \widetilde{\Delta}_{LL}(t)
\end{equation}
where $\widetilde{\Delta}_{LL}(t)$ includes contributions from the $\eta'$-$\pi$ state and from the 
ground state scalar meson, and will be modelled by the bubble sum depicted in Figure \ref{fig:bubblesum}. 
For the lighter quark masses, the $\eta'$-$\pi$ state becomes important, and it is more difficult to 
disentangle the $a_0$ and $a_0^*$ contributions. In the fits to the local-local propagator described 
in the next section, we have neglected the quark mass dependence of the excited state $a_0^*$ term and 
used the estimate (\ref{eq:heavy}) for all quark masses. 

By resumming the multiple bubble and scalar propagator graphs shown in Figure \ref{fig:bubblesum}, 
the full propagator may be written in terms of the one-loop bubble function $B(p)$ defined in 
equation (\ref{eq:B}),
\begin{equation}
\label{eq:bubblesum}
\Delta(p) = 
    -4r_0^2 B^{ren}
    + 32r_0^2 \left(f_s - f_s (2m_{\pi}^2/f^2) B^{ren} \right) 
   \left(f_s - f_s (2m_{\pi}^2/f^2) B^{ren}\right) P_{a0} \;
\end{equation}
where $f_s$ is the coupling of the scalar meson to the $\bar{\psi}\psi$ density, $B^{ren}(p)$ is the resummed multiple-bubble function,
\begin{equation}
B^{ren} = \frac{B(p)}{1+(2m_{\pi}^2/f^2)B}
\end{equation} 
and $P_{a0}$ is the resummed $a_0$ propagator,
\begin{equation}
P_{a0}^{-1} = \hat{p}^2 + m_s^2 + 32(f_sm_{\pi}^2/f^2)^2B^{ren}
\end{equation}
Note that the negative sign associated with the hairpin insertion 
has been taken outside of the bubble function, i.e. the functions $B(p)$ and $B^{ren}(p)$
are both positive. 

We use the bubble function $B(p)$ defined for the lattice version 
of the full chiral Lagrangian with the same lattice spacing as 
the quark lattice.   In (20,21), the lattice 
four-momentum is defined by $\hat{p}\equiv 2\sin(p/2)$ as appropriate for bosonic
propagators.
If we are to use the pion masses in Table I, which were determined from fitting 
the t-space pion propagators, then the pion 
masses used in lattice propagators above are given by $\tilde{m}_{\pi}=2\sinh(m_{\pi}/2)$.

This analysis includes all contributions to the scalar 
propagator which can be computed by direct 
$\eta'$-$\pi$ bubble summation and mixing with the $a_0$
scalar meson state.   We use these expressions in our analysis 
of the lattice data on the isovector, scalar propagator.  

The quenched theory does introduce additional contributions 
which could modify the behavior of the bubble terms.  
For example, pions can interact through double hairpin 
exchange diagrams whose infrared behavior is doubly 
enhanced.   Therefore, virtual $\eta '$ processes do not 
decouple in the infrared limit as is normally expected for 
soft pion interactions but can generate long range 
forces between mesons in the quenched theory.  Preliminary 
estimates of these effects indicate they are negligible 
for the range of pion masses and lattice volume 
considered in this paper and they have not been included in our analysis.

\section{Global Fits and Determination of Chiral 
Lagrangian Parameters}

In Section 2 we showed that the data for the scalar propagator for heavier quark masses was dominated 
by the scalar meson $a_0$ intermediate state (after removing the excited $a_0^*$ contribution, as 
discussed in that Section), while for light masses, it is dominated by the $\eta'$-$\pi$ intermediate 
state. To get a consistent fit for all masses, we have found that the resummed formula (\ref{eq:bubblesum}) 
(Figure \ref{fig:bubblesum}) produces the most stable results. (The simpler alternative of using the sum 
of a scalar meson pole plus the single bubble function gives fits with only slightly worse $\chi^2$'s 
but the resulting fit parameters $f_s$ and $m_s$ are less stable as a function of quark mass.) 
All of the fits discussed in this section are based on the formula (\ref{eq:bubblesum}) with the parameters
$r_0$ and $f$ fixed at their previously determined values (\ref{eq:fixedparams}). We present the results of 
two different fitting procedures. In the first set of fits (Table I), 
we take the hairpin mass insertion parameter $m_0$ to be
fixed at the value (\ref{eq:fixedparams}) and use the scalar propagator fits to extract the scalar 
meson Lagrangian parameters $f_s$ and $m_s$. In the second set of fits (Table II), 
we let $m_0$ be a fit paramter
and investigate how well the scalar propagator data by itself determines the value of the hairpin insertion.

\begin{table}
\centering
\caption{ Fit parameters for the scalar propagator, with $m_0$ fixed. 
The fit range is $t=[1-6]$.}
\label{tab:masses}
\begin{tabular}{||c|c|c|c|c||}
\hline 
$\kappa$ & $m_P$ & $f_s$ & 
$m_s$ & $\chi^2/dof $ \\
\hline
.1400 & .603(2) & .0544(10) & 1.20(2) & 1.7/4 \\
.1405 & .556(2) & .0548(10) & 1.19(2) & 2.0/4 \\
.1410 & .505(2) & .0550(11) & 1.17(2) & 1.9/4 \\
.1415 & .450(3) & .0554(12) & 1.16(3) & 1.7/4  \\
.1420 & .386(3) & .0559(14) & 1.16(4) & 1.7/4 \\
.1423 & .342(4) & .0565(19) & 1.18(5) & 2.3/4 \\
.1425 & .307(4) & .0569(25) & 1.19(7) & 3.7/4 \\
.1427 & .267(5) & .0568(38) & 1.20(12) & 5.6/4 \\
.1428 & .245(6) & .0571(54) & 1.29(18) & 6.2/4 \\
\hline
\end{tabular}
\end{table}

The fits which extract the scalar parameters, with $m_0$ held fixed, are given in
Table \ref{tab:masses}. We see that the value of the ground state scalar mass parameter $m_s$
is rather well determined 
for the heaviest quark masses, but becomes less accurate for lighter masses, where the $\eta'$-$\pi$ state 
dominates. A linear fit to the scalar mass values in Table \ref{tab:masses} gives
a value of 
\begin{equation}
m_s = 1.14(7) 
\end{equation}
in the chiral limit. Using the charmonium scale $a^{-1}=1.18$ GeV, this gives
a scalar meson mass of $1.34(9)$ GeV. The scalar decay constant $f_s$ in the chiral 
limit is
\begin{equation}
\label{eq:fs}
f_s = .057(3)
\end{equation}

\begin{table}
\centering
\caption{ Fit parameters for the scalar propagator, with $m_s$ fixed to values shown. 
The fit range is $t=[1-6]$.}
\label{tab:m0fits}
\begin{tabular}{||c|c|c|c|c||}
\hline 
$\kappa$ & $m_s$ & $f_s$ & 
$m_0$ & $\chi^2/dof $ \\
\hline
.1400 & 1.200 & .0568(12) & .336(22) & 2.1/4 \\
.1405 & 1.185 & .0570(12) & .335(25) & 2.3/4 \\
.1410 & 1.170 & .0571(13) & .335(27) & 2.3/4 \\
.1415 & 1.156 & .0574(13) & .338(30) & 2.1/4 \\
.1420 & 1.142 & .0575(14) & .346(33) & 2.3/4 \\
.1423 & 1.133 & .0578(15) & .358(36) & 3.0/4 \\
.1425 & 1.128 & .0576(16) & .360(43) & 4.5/4 \\
.1427 & 1.122 & .0564(18) & .332(51) & 6.4/4 \\
.1428 & 1.119 & .0555(20) & .298(64) & 6.7/4 \\
\hline
\end{tabular}
\end{table}

\begin{figure}
\vspace*{2.0cm}
\includegraphics{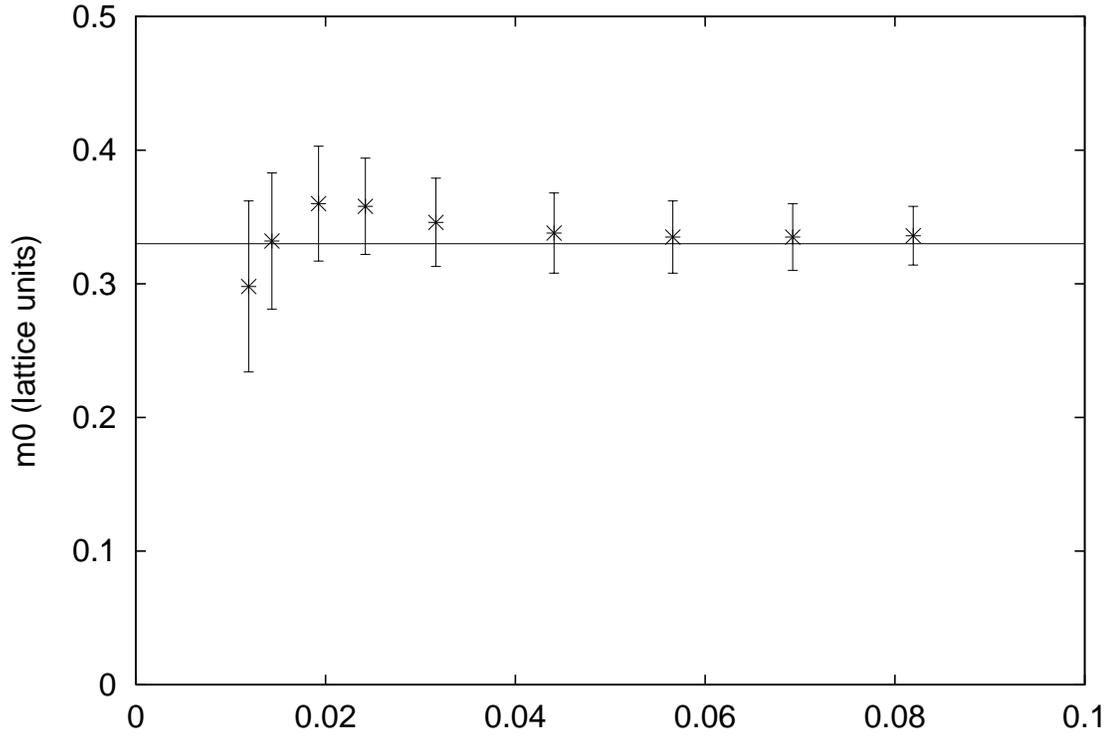}
\vspace{11.5cm}
\caption[]{Results for the $\eta'$ hairpin mass insertion extracted from fits to the scalar 
propagator. The solid line is the value obtained from direct analysis of the hairpin diagram
in Reference \cite{chlogs} (after removing a $\sqrt{3}$ flavor factor included in the 
definition of $m_0$ in that reference).  }
\label{fig:m0}
\end{figure}

In addition to extracting scalar parameters,
it is interesting to carry out fits to the scalar propagator with $m_0$ as a fit parameter. 
This procedure emphasizes the fact that the scalar propagator data alone provides a fairly accurate 
estimate of $m_0$ which is independent of, and consistent with, the previous estimates in 
Reference \cite{chlogs}. As we saw in the previous fits, the scalar mass $m_s$ is well-determined 
for the heaviest quarks, but is poorly determined at the light-quark end. To obtain stable fits with 
$m_0$ as a fit parameter, we fix $m_s$ to be given by a linear fit obtained from the 
three heaviest quark values, $m_s = 1.106 + 1.145 m_q$. 
where $m_q$ is the bare quark mass $(\kappa^{-1}-\kappa_c^{-1})/2$ in lattice units.
In this set of fits, $f_s$ is found to be approximately independent of 
quark mass and is well-approximated by (\ref{eq:fs}) for all quark masses. The values of the $\eta'$
mass parameter $m_0$ obtained from these fits are given in Table \ref{tab:m0fits} and
plotted in Figure \ref{fig:m0}. A linear fit to this data gives
\begin{equation}
m_0 = 0.34(4)
\end{equation}
in the chiral limit.
This result, obtained from the scalar propagator data alone, is quite consistent with our previous estimates 
which included  direct $\eta'$ hairpin measurements, a topological susceptibility calculation, and QCL effects 
in the pseudoscalar channel \cite{chlogs}. 

\begin{figure}
\vspace*{2.0cm}
\includegraphics{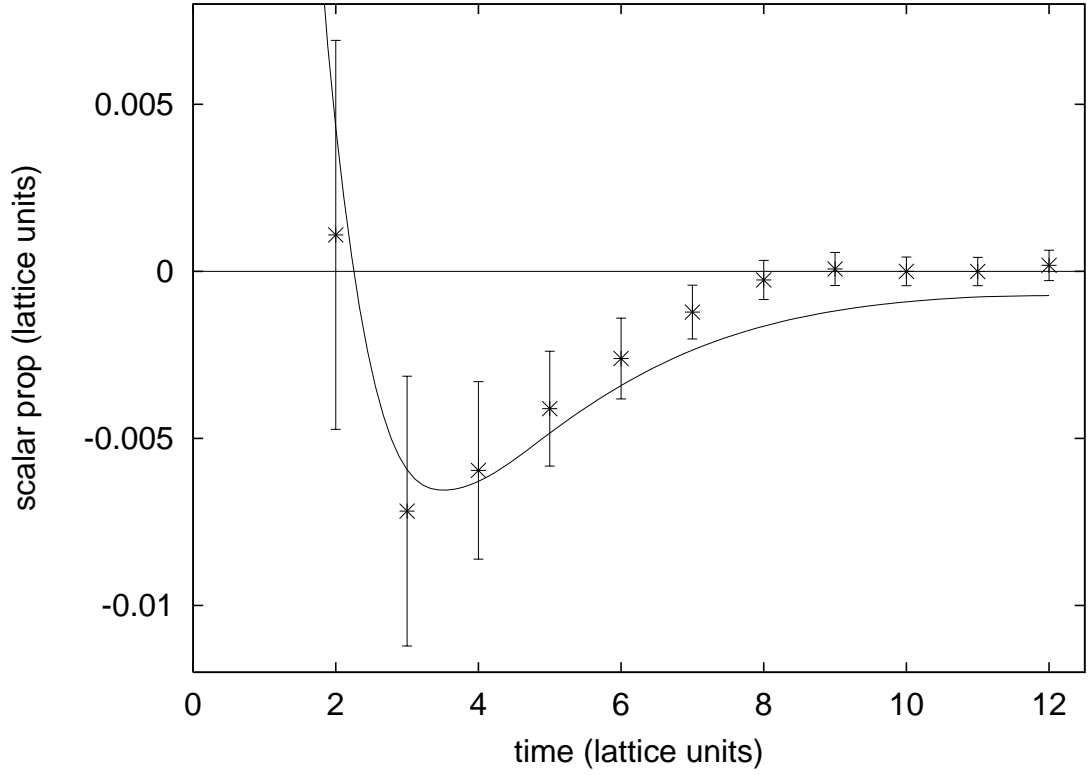}
\vspace{11.5cm}
\caption[]{Comparison of scalar propagator for $\kappa=.1427$ with
the bubble sum formula (\ref{eq:bubblesum}) fitted to the interval t=1-6. 
}
\label{fig:tail_light}
\end{figure}

One feature of the scalar propagator which appears to be somewhat inconsistent 
with our theoretical model is the behavior at  larger time separations,
$t\geq 7$. For the lightest quarks, we saw that the propagator was well-described in the region $t=3$ to 6 by the one-loop $\eta'$-$\pi$ bubble. However,
beyond $t=6$, the propagator appears to go to zero (from below) more rapidly than the theoretical curve 
predicted by either the one-loop calculation or
the bubble-sum formula, as shown for $\kappa=.1427$ in Figure \ref{fig:tail_light}. 
If it is a significant departure from the chiral symmetry prediction, it could be an indication of strong $\eta'$-$\pi$ 
interactions near threshold. 
Such interactions would normally be ruled out by soft-pion arguments, but may 
arise due either to explicit chiral symmetry breaking of the Wilson-Dirac action or because of higher order 
quenched effects.  
For the heaviest quarks, we also find that the propagator appears to have 
an additional positive contribution for $t\geq 7$. 
For example, we can obtain a good fit 
(to both the heavy and light mass data at large time) by adding a 
small $\pi$ propagator 
term to the fit model.  However, the origin of such a contribution 
is not presently understood within our model. 

\section{Discussion}

	We have presented convincing evidence that the 
quenched scalar, isovector meson propagator exhibits 
a prominent ``quenched chiral loop'' (QCL) artifact at 
light quark mass which is properly interpreted 
as the effect of an $\eta'$-$\pi$ intermediate state.  QCL effects that have 
been observed in other quantities \cite{chlogs,CPPACS} 
have been rather subtle, involving relatively small 
deviations from full QCD chiral behavior.   By contrast, 
the QCL effect discussed here is an  unmistakable chiral
power singularity, and 
in fact dominates the scalar correlator for the lightest quark masses we 
have studied.   The negative spectral weight of this 
contribution is a particularly vivid illustration of the 
oft-stated truism that the quenched approximation 
violates unitarity.

The results presented here also provide further 
evidence of the effectiveness of the pole-shifting 
ansatz of the modified quenched approximation. It is 
only after pole-shifting that we are able to study 
the lightest quark masses for which the 
$\eta'$-$\pi$ state becomes obvious. It should be 
remarked, however, that even for the heavier 
quark masses, the QCL effect is not negligible, and 
must be included in the fitting function in  order to 
draw correct conclusions about the scalar meson 
mass. For example, fitting the scalar propagator  at the heavier quark  masses 
with a simple exponential and ignoring the $\eta'$-$\pi$ 
contribution would lead to the erroneous conclusion 
that the scalar meson becomes {\em heavier} as the quark 
mass gets lighter. The resulting effect on the chiral 
extrapolation is at least as severe as typical QCL 
effects in the quenched light hadron spectrum. For 
pion masses below about 400 MeV, the 
$\eta'$-$\pi$ contribution becomes dominant, making 
the scalar meson mass difficult to determine 
accurately.  Nevertheless, by combining lattice 
calculations with quenched chiral perturbation 
theory, more extensive studies of scalar meson 
spectroscopy in the quenched approximation 
should be feasible. Finally, the prominence of the 
QCL effect in the scalar valence propagator suggests 
that this propagator be added to the list of standard 
bellwether quantities (such as topological 
susceptibility, $\eta'$ mass, and long-range static potential) 
which are expected to be particularly sensitive to 
the difference between quenched and full QCD gauge 
ensembles.

\section{Acknowledgements}

The work of W. Bardeen and E. Eichten was performed 
at the Fermi National Accelerator Laboratory, which is 
operated by University Research Association,
Inc., under contract DE-AC02-76CHO3000. 
The work of A. Duncan was supported in part by 
NSF grant PHY00-88946.
The work of N. Isgur was supported by DOE Contract DE-AC05-84ER40150
under which the Southeastern Universities Research Association (SURA)
operates the Thomas Jefferson National Accelerator Facility (Jefferson Lab). 
The work of H. Thacker was supported in part by the
 Department of Energy under grant DE-FG02-97ER41027.

\end{document}